\documentclass[aps,prb,reprint,twocolumn,longbibliography]{revtex4-2}
\pdfoutput=1

\usepackage{graphicx}
\usepackage{amsmath,amsfonts,amssymb}
\usepackage{color}
\usepackage{hyperref}

\begin{document}

\title{Existence of Hebel-Slichter peak in unconventional kagome superconductors}

\author{Yi Dai}
\affiliation{Niels Bohr Institute, University of Copenhagen, DK-2200 Copenhagen, Denmark}

\author{Andreas Kreisel}
\affiliation{Niels Bohr Institute, University of Copenhagen, DK-2200 Copenhagen, Denmark}

\author{Brian M. Andersen}
\affiliation{Niels Bohr Institute, University of Copenhagen, DK-2200 Copenhagen, Denmark}

\date{October 17, 2024}

\begin{abstract}
We perform a theoretical investigation of the spin susceptibility of unconventional superconductivity on the kagome lattice. Despite the existence of a sign-changing gap structure, which sums to zero over the Fermi surface, we show that such unconventional pairing states may exhibit a Hebel-Slichter peak in the temperature-dependent spin-lattice relaxation rate. It originates from destructive sublattice interference effects. For the same reason, unconventional pairing states on the kagome lattice tend not to exhibit a neutron resonance peak. These results supplement previous theoretical studies of the surprising robustness of sign-changing gap structures to disorder on the kagome lattice. Taken together these findings imply that unconventional superconductivity on the kagome lattice is deceptive in the sense that its properties may appear similar to conventional non-sign-changing superconductivity.
These results may be of relevance to the superconducting state of the kagome superconductors $A$V$_3$Sb$_5$ ($A$: K, Rb, Cs) and CsTi$_3$Bi$_5$.
\end{abstract}

\maketitle

\section{Introduction}
For progress in the understanding of unconventional superconductivity it is crucial to experimentally determine the nature of the superconducting ground state. This is a demanding collective task that requires agreement between a vast range of different experimental probes. Pinpointing the pairing state includes determination of the spin structure of the Cooper pairs and their relative spatial dependence, i.e. the realized irreducible representation of the associated crystal point group~\cite{Volovik1985,SigristUeda91,Annett90}. Together these properties hold valuable clues to the dominant fluctuations driving the Cooper pairing~\cite{Scalapino2012,Chubukov2012Pairing,Kreisel_review,KreiselOS,Romer2019}. In this endeavor, it is important to identify any possible sign changes of the gap function, which is experimentally challenging as most probes are insensitive to the phase of the superconducting order parameter. An exception is scattering off impurities or sample edges which does allow for access to the phase. Likewise, two-particle correlation functions can be phase sensitive as evidenced for example by the spin susceptibility featuring a Hebel-Slichter peak in the nuclear magnetic resonance (NMR) spin-lattice relaxation rate, and the neutron resonance peak detected by inelastic neutron scattering experiments. We return to a discussion of these signatures further below, demonstrating that for the kagome lattice the phase-sensitiveness is wiped out by destructive sublattice interference effects.    

In recent years, the discussion of possible unconventional superconductivity on the kagome lattice has been motivated by the discovery of superconductivity in vanadium-based kagome metals $A$V$_3$Sb$_5$ ($A$: K, Rb, Cs)~\cite{Ortiz2020CsV3Sb5,OrtizEA21,YinEA21}. The kagome lattice is particularly interesting since the associated band structure exhibits a flat band, van Hove singularities, and Dirac points, as seen in Fig.~\ref{fig1}. Importantly, the distribution of sublattice weights of the eigenstates on the Fermi surface, also illustrated in Fig.~\ref{fig1}c, plays an important role in determining the leading instabilities arising from interactions~\cite{KieselEA12,Kiesel2013Unconventional,WuEA22,Scammell2023}. This may  be relevant for the $A$V$_3$Sb$_5$ materials where superconductivity appears in proximity to a charge-density wave phase~\cite{Ortiz2019New,Kenney2021Absence,Jiang2021Unconventional,Chen2021Roton,Zhao2021Cascade,ParkEA21,LinEA22,Denner2021,Tazai2022mechanism,Christensen2021,Ferrari2022,Christensen2022}. For the superconducting phase, theoretical studies have explored Cooper pairing arising both from purely electronic fluctuations~\cite{Yu2012,KieselEA12,Wang2013,ParkEA21,Wu2021Nature,Tazai2022mechanism,Wen2022superconducting,RomerEA22,Wu2021Nature,He2022Strong-coupling,LinEA22,Bai2022effective,Profe2024} and via the importance of phonon contributions~\cite{Wu2EA22,TanEA21,Zhang2021firstprinciples,Zhong2022Testing,Wang_phonon_2023,Ritz23}. From pairing via spin- and charge-fluctuations, the $E_2$ irreducible representation with $d$-wave (or $d\pm id$) pairing symmetry stands out as a leading candidate~\cite{RomerEA22}.

Experimentally, the nature of the superconducting ground state in the $A$V$_3$Sb$_5$ compounds remains unresolved at present~\cite{Wilsonreview}, with conflicting evidence for both standard nodeless non-sign-changing gaps and nodal unconventional superconducting order~\cite{Chen2021Roton,Liang2021Three-dimensional,Xu2021Multiband,Zhao2021nodal,Guguchia2022Tunable,Mielke2022Time-reversal}. A Knight shift suppression and the existence of a Hebel-Slichter peak below $T_c$ in the spin-lattice relaxation rate measured by NMR experiments were interpreted as evidence for $s$-wave spin-singlet superconductivity~\cite{Mu2021S-wave}.  This agrees with recent laser ARPES measurements reporting isotropic (momentum-independent) spectroscopic gaps~\cite{ZhongEA23}. Penetration depth data and specific heat measurements on CsV$_3$Sb$_5$ have been analysed in terms of an anisotropic, but non-sign-changing gap with a finite small minimum gap~\cite{OrtizEA21,Duan2021Nodeless,Gupta2022Microscopic,RoppongiEA22,Gupta2022Two}. Reference~\onlinecite{RoppongiEA22} measured electron irradiation effects on the penetration depth and found no evidence for disorder-generated low-energy density of states enhancements, as expected from sign-changing gap functions. Likewise, a non-sign-changing gap function appears consistent with the absence of in-gap bound states near nonmagnetic impurities~\cite{Xu2021Multiband} and a weak dependence of the critical transition temperature $T_c$ on the residual resistivity ratio (sample quality)~\cite{Zhang2023}.

Indeed, the response of superconductivity to disorder can play an important role in determining the nature of superconductivity since it typically acts as a phase-sensitive probe~\cite{Balatsky_review,Gastiasoro2016,Gastiasoro2018,RomerRaising,Andersen2023}. Disorder effects can be studied both from the possible generation of impurity bound states detectable by local probes and an overall disorder-averaged response verified by thermodynamic probes or transport measurements. When nonmagnetic disorder generates in-gap bound states or if it strongly reduces $T_c$, it is traditionally a strong indicator of an unconventional superconducting condensate. These conclusions about the effects of disorder on unconventional sign-changing gap structures are, however, partly invalidated on the kagome lattice. As recently demonstrated, the sublattice weight of the band eigenstates play a crucial role for determining the response of superconductivity to disorder on the kagome lattice~\cite{Sofie2023}. Specifically, even though the unconventional gap function averages to zero over the Fermi surface, $\sum_{\bf k}\Delta({\bf{k}})=0$, atomic-scale disorder only scatters to a limited allowed region of the Fermi surface. For even-parity (spin-singlet) superconductivity this has important consequences, including the absence of in-gap bound states and a slow $T_c$-suppression rate~\cite{Sofie2023}. In this light, unconventional (spin-singlet) superconductivity on the kagome lattice is protected from nonmagnetic disorder, a fact that may be of importance for the interpretation of several of the experiments mentioned above~\cite{Xu2021Multiband,RoppongiEA22,Zhang2023}.

In this paper we study the spin susceptibility in the superconducting state on the kagome lattice within a minimal tight-binding model. Based on the evidence for spin-singlet Cooper pairs, we focus on the spin-lattice relaxation rate in the even-parity states: $A_1$ ($s$-wave) and $E_2$ ($d$-wave) superconducting order. Surprisingly, despite a fully compensated sign-changing gap structure, the relaxation rate exhibits a pronounced Hebel-Slichter peak upon entering the superconducting state. In addition, the neutron resonance peak expected for sign-changing gap structures is largely wiped out~\cite{ROSSATMIGNOD199186,Mook1993,Christianson2008,Inosov2010,StockNR}. The origin of these effects can be traced to the peculiar properties of the matrix elements from sublattice to band space on the kagome lattice. Thus, in short, unconventional spin-singlet superconductivity on the kagome lattice takes the deceptive appearance of a conventional superconductor. This conclusion is in line with the unusual disorder effects mentioned above~\cite{Sofie2023}. Our results lead to a reconsideration of the superconducting state of kagome superconductors under current investigations, and specifically invalidates the conclusion that a Hebel-Slichter peak implies conventional $s$-wave superconductivity for these compounds. 

\section{Basic band structure of the kagome lattice}\label{sec:kagome_lattice}
\begin{figure}[tb]
    \includegraphics[width=\linewidth]{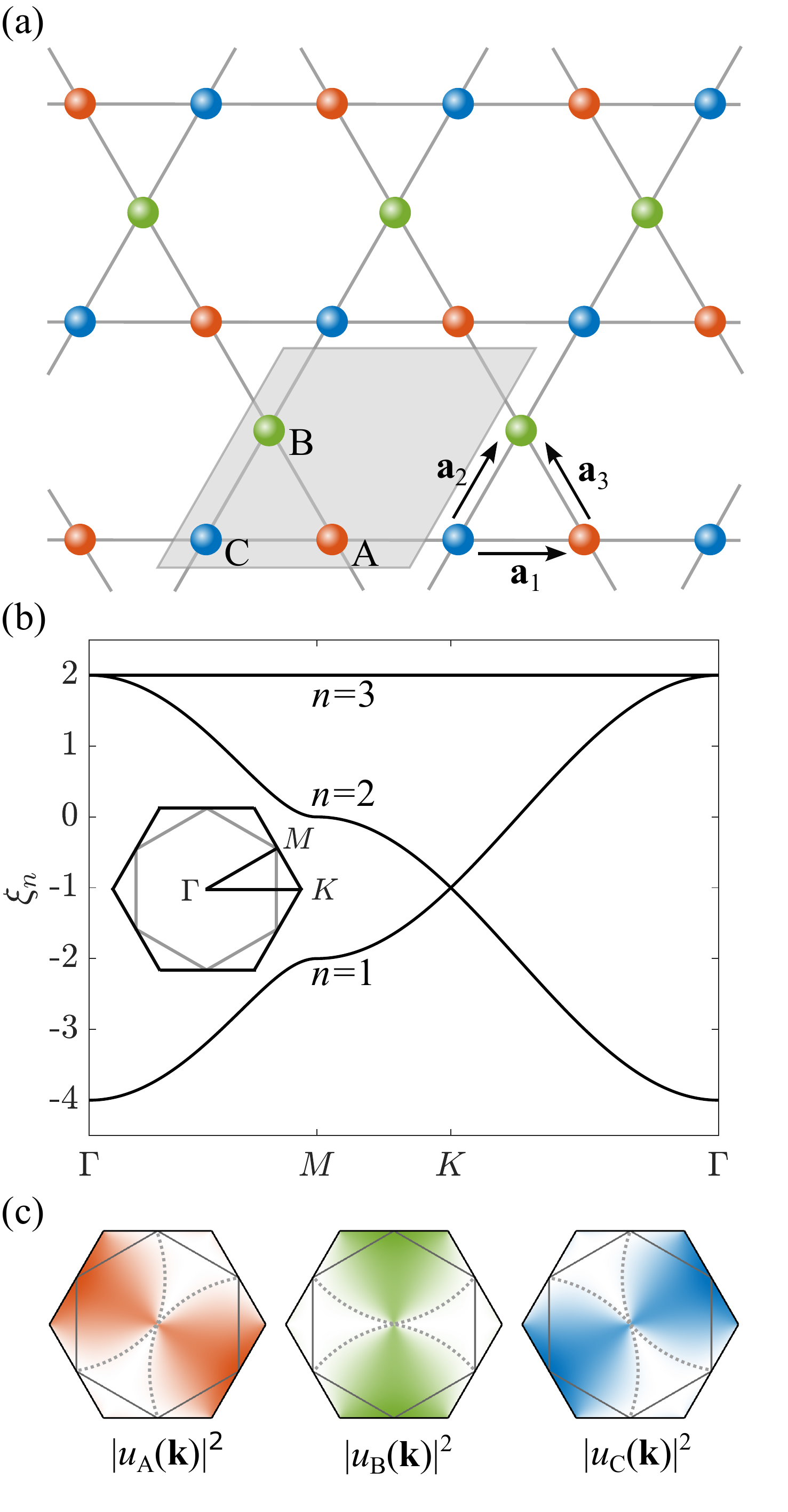}
    \caption{(a) Illustration of the kagome lattice with sublattice vectors $\mathbf{a}_n$. (b) Energy bands of the tight-binding model   $\xi_n(\mathbf{k})$ in units of $t$ along the path shown in the inset. The dark gray lines inside in the Brillouin zone (BZ) define the Fermi surface at $\mu = 0$. (c) Distribution of the sublattice weights across the BZ for the middle band ($n$ = 2). The colors refer to the same sublattices shown in (a). The dashed lines sketch the area where the sublattice weights are zero.}
    \label{fig1} 
\end{figure}
The minimal tight-binding model of electronic states on the kagome lattice is shown in Fig.~\ref{fig1}. The Hamiltonian of the tight-binding model is given by
\begin{equation}
\label{eq_1_1}
    \begin{split}
    \mathcal{H}_0 = \sum_{\mathbf{k},\sigma}\psi^{\dagger}_{\mathbf{k}  \sigma}H_0 (\mathbf{k})\psi_{\mathbf{k} \sigma},
    \end{split}
\end{equation}
where $\psi_{\mathbf{k} \sigma} = \begin{pmatrix}
        c_{\mathbf{k} \sigma A} & c_{\mathbf{k} \sigma B} & c_{\mathbf{k} \sigma C}
    \end{pmatrix}^T$ and
\begin{equation}
\label{eq1_2}
    \begin{split}
    H_0 (\mathbf{k}) = -\begin{pmatrix}
       \mu & t\cos k_{3} & t\cos k_{1}\\
       t\cos k_{3} & \mu & t\cos k_{2}\\
       t\cos k_{1} & t\cos k_{2} & \mu
    \end{pmatrix}.
    \end{split}
\end{equation}
Here $k_n = \mathbf{k} \cdot \mathbf{a}_n$, where $\mathbf{a}_1 = \frac{1}{2}\begin{pmatrix} 1&0\end{pmatrix}$, $\mathbf{a}_2 = \frac{1}{2}\begin{pmatrix} \frac{1}{2}&\frac{\sqrt{3}}{2}\end{pmatrix}$ and $\mathbf{a}_3 = \frac{1}{2}\begin{pmatrix} -\frac{1}{2}&\frac{\sqrt{3}}{2}\end{pmatrix}$. $\mu$ refers to the chemical potential, we use the NN hopping integrals $t=1$ as energy unit in the following.
The Hamiltonian is diagonalized by a unitary transformation, $u_{n \alpha }^{\ast}(\mathbf{k}) H_{0,\alpha\beta}(\mathbf{k}) u_{\beta m}(\mathbf{k})=\xi_n(\mathbf{k}) \delta_{nm}$ yielding the band energies $\xi_n(\mathbf{k})$ and the eigenstates $u_{n \alpha}(\mathbf{k})$ of band $n$.
The resulting band structure is shown in Fig.~\ref{fig1}(b) and features a Dirac point at the K point and two van Hove singularities at the M point. In addition, for the nearest-neighbor (NN) tight-binding model there is a flat band that acquires  dispersion upon including further neighbor hoppings. The kagome lattice is endowed with a property which has become known as sublattice interference~\cite{KieselEA12} for which specific hopping trajectories interfere destructively and result in electronic wavefunctions that localize on specific sites inside the unit cell. Specifically, electronic states at the upper van Hove singularity, at $\mu=0$, are localized on only one of the three sublattice sites in the unit cell, as illustrated in Fig.~\ref{fig1}(c). By contrast, states at the lower van Hove singularity near $\mu=-2$ localize on two of the three sublattice sites. 

\section{Spin-lattice relaxation rate}\label{sec:kagome_SC}

\subsection{Revisiting the one-band square lattice}

In order to set the stage for the discussion of the superconducting spin susceptibility on the kagome lattice, we start by briefly revisiting basic properties of the spin susceptibility in the superconducting state on the square lattice. In that case, the bare retarded BCS spin susceptibility $\chi^{+-}_0(\mathbf{q},\omega)$ at momentum $\mathbf{q}$ and frequency $\omega$ is given by 
\begin{widetext}
\begin{align}
   \chi^{+-}_0(\mathbf{q},\omega)  =
    \frac{1}{\mathcal{N}} \! \sum_{\mathbf{k}, E>0}\!&\left[ \left(1 \! - \! \frac{\xi_{\mathbf{k}}\xi_{\mathbf{k} \! + \! \mathbf{q}} \! + \! \Delta^*_{\mathbf{k} \! + \! \mathbf{q}}\Delta_{\mathbf{k}}}{E_{\mathbf{k}}E_{\mathbf{k} \! + \! \mathbf{q}}}\right)  \! \frac{1 \! - \! f(E_{\mathbf{k}}) \! - \! f(E_{\mathbf{k} \! + \! \mathbf{q}})}{\omega \! + \! E_{\mathbf{k} \! + \! \mathbf{q}} \! + \! E_{\mathbf{k}} \! + \! i\eta}\right.
     \! + \!  \left(1 \! - \! \frac{\xi_{\mathbf{k}}\xi_{\mathbf{k} \! + \! \mathbf{q}} \! + \! \Delta^*_{\mathbf{k} \! + \! \mathbf{q}}\Delta_{\mathbf{k}}}{E_{\mathbf{k}}E_{\mathbf{k} \! + \! \mathbf{q}}}\right) \! \frac{f(E_{\mathbf{k}}) \! + \! f(E_{\mathbf{k} \! + \! \mathbf{q}}) \! - \! 1}{\omega \! - \! E_{\mathbf{k} \! + \! \mathbf{q}} \! - \! E_{\mathbf{k}} \! + \! i\eta}
     \notag\\
    & \! \! + \! \!   \left(1\! + \!\frac{\xi_{\mathbf{k}}\xi_{\mathbf{k} \! + \! \mathbf{q}}\! + \!\Delta^*_{\mathbf{k} \! + \! \mathbf{q}}\Delta_{\mathbf{k}}}{E_{\mathbf{k}}E_{\mathbf{k} \! + \! \mathbf{q}}}\right)\frac{f(E_{\mathbf{k}})\! - \!f(E_{\mathbf{k} \! + \! \mathbf{q}})}{\omega\! + \!E_{\mathbf{k} \! + \! \mathbf{q}}\! - \!E_{\mathbf{k}}\! + \!i\eta}
    \! + \!  \left. \left(1\! + \!\frac{\xi_{\mathbf{k}}\xi_{\mathbf{k} \! + \! \mathbf{q}}\! + \!\Delta^*_{\mathbf{k} \! + \! \mathbf{q}}\Delta_{\mathbf{k}}}{E_{\mathbf{k}}E_{\mathbf{k} \! + \! \mathbf{q}}}\right)\frac{f(E_{\mathbf{k} \! + \! \mathbf{q}})\! - \!f(E_{k})}{\omega\! + \!E_{\mathbf{k}}\! - \!E_{\mathbf{k} \! + \! \mathbf{q}}\! + \!i\eta}\right],
    \label{eq_2_1}
\end{align}
\end{widetext}
where $\mathcal{N}$ denotes the number of points summed over in the Brillouin zone (BZ), and $\eta$ is an infinitesimal positive factor arising from the analytical continuation. Additionally, $\xi_\mathbf{k}$ is the electron dispersion, $\Delta_\mathbf{k}$ the superconducting order parameter, $E_\mathbf{k} = \sqrt{\xi_\mathbf{k}^2 + |\Delta_\mathbf{k}|^2}$ denotes the energy of superconducting quasiparticles, and $f(E_\mathbf{k})$ is the Fermi-Dirac distribution function. Because of the Nambu particle-hole symmetry, we can sum only over positive $E_\mathbf{k}$.

\begin{figure}[tb]
    \includegraphics[width=\linewidth]{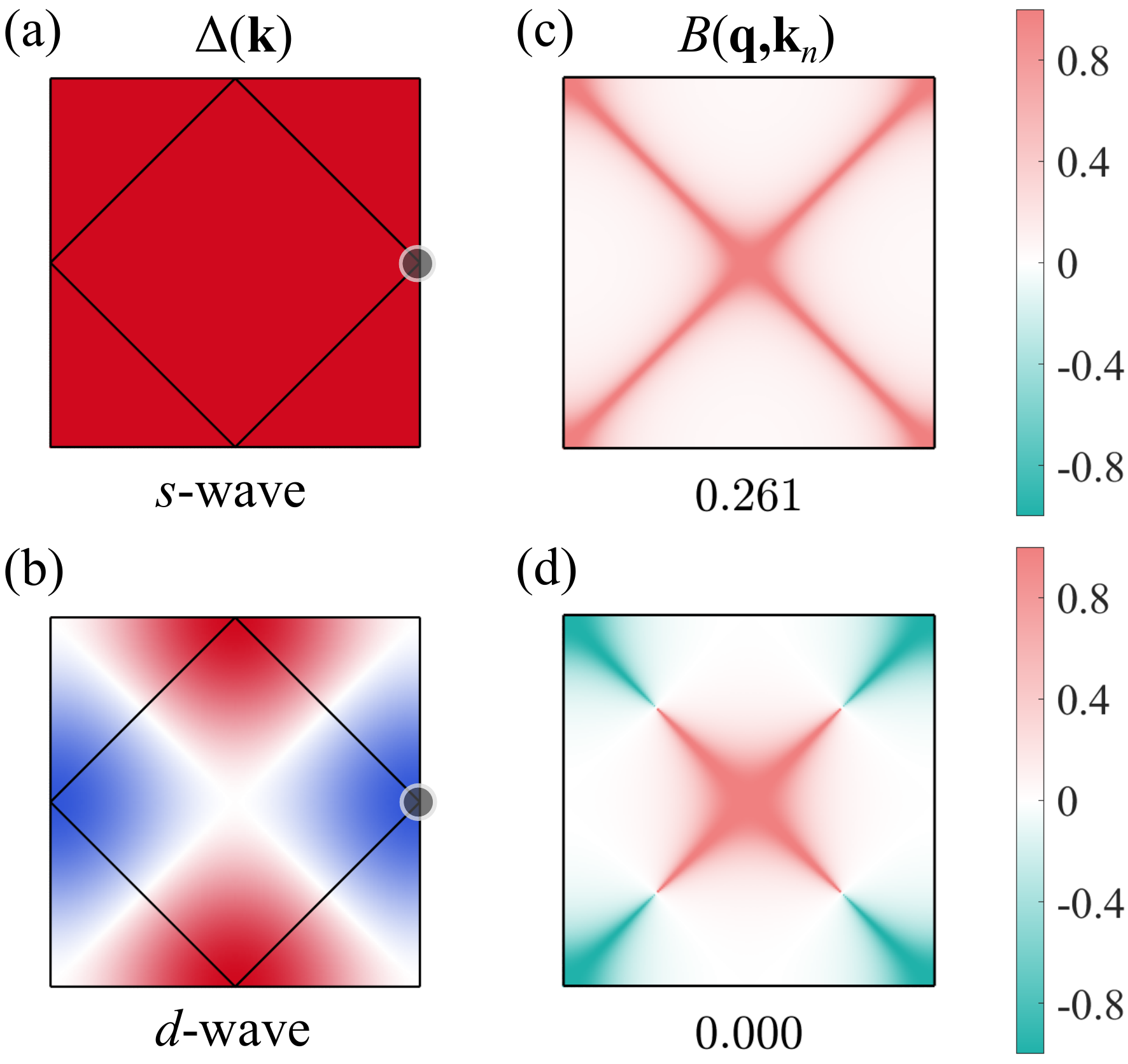}
    \caption{Panels (a) and (b) display the $s$-wave and $d_{x^2-y^2}$-wave order parameters in $\mathbf{k}$-space in the first BZ for the square lattice. The black lines indicate the Fermi surface at $\mu = 0$. Panels (c) and (d) display $B(\mathbf{q},\mathbf{k}_n)$ defined in Eq.~\eqref{eq2_21} in $\mathbf{q}$-space for the $s$-wave and $d_{x^2-y^2}$-wave order parameters, respectively. The chosen momentum point $\mathbf{k}_n$ is shown by the black dot in (a) and (b). The numbers below (c) and (d) indicate the sum $\sum_{\mathbf{q}}B(\mathbf{q},\mathbf{k}_n)$.}
    \label{fig_sq_cancellation}
\end{figure}
The spin-lattice relaxation rate is related to the imaginary part of spin susceptibility $\chi^{+-}_0(\mathbf{q},\omega)$ by \cite{Moriya1963Effect,Franziska2012Magnetism}
\begin{equation}
\label{eq_2_2}
    \begin{split}
    \alpha \equiv \frac{1}{T_1T} \propto \lim_{\omega\rightarrow 0} \frac{1}{\mathcal{N}}\sum_{\mathbf{q}}\mathop{\text{Im}}\frac{\chi^{+-}_0(\mathbf{q},\omega)}{\omega}.\\
    \end{split}
\end{equation}
The tight-binding model of the square lattice is 
\begin{equation}
\label{eq_2_3}
    \begin{split}
    \mathcal{H}_0 = \sum_{\mathbf{k}\sigma}\left[-2t\left(\cos k_x+\cos k_y\right)-\mu\right]c^{\dagger}_{\mathbf{k}\sigma}c_{\mathbf{k}\sigma}.
    \end{split}
\end{equation}
For the purpose of illustration, we consider two distinct superconducting cases: (1) a conventional isotropic $s$-wave order parameter $\Delta_{\mathbf{k}}^s = \Delta_0 = 0.2$, and (2) a sign-changing $d$-wave order parameter given by $\Delta ^d _\mathbf{k} = \frac{\Delta_0}{2} \left(\cos k_x-\cos k_y \right)$.
In Eq.~\eqref{eq_2_1}, only the last two terms can give non-zero values for the imaginary part in the superconducting states in the limit $\omega\rightarrow 0$~\cite{Tinkham75}. In addition to a sharp onset of the order parameter at $T_c$, the factors in the associated brackets 
\begin{equation}
    \left(1\! + \!\frac{\xi_{\mathbf{k}}\xi_{\mathbf{k} \! + \! \mathbf{q}}\! + \!\Delta^*_{\mathbf{k} \! + \! \mathbf{q}}\Delta_{\mathbf{k}}}{E_{\mathbf{k}}E_{\mathbf{k} \! + \! \mathbf{q}}}\right),
\end{equation}
are key to the existence of a Hebel-Slichter peak. We can illustrate this by defining a factor
\begin{figure}[tb]
     \centering
    \includegraphics[width=.88\linewidth]{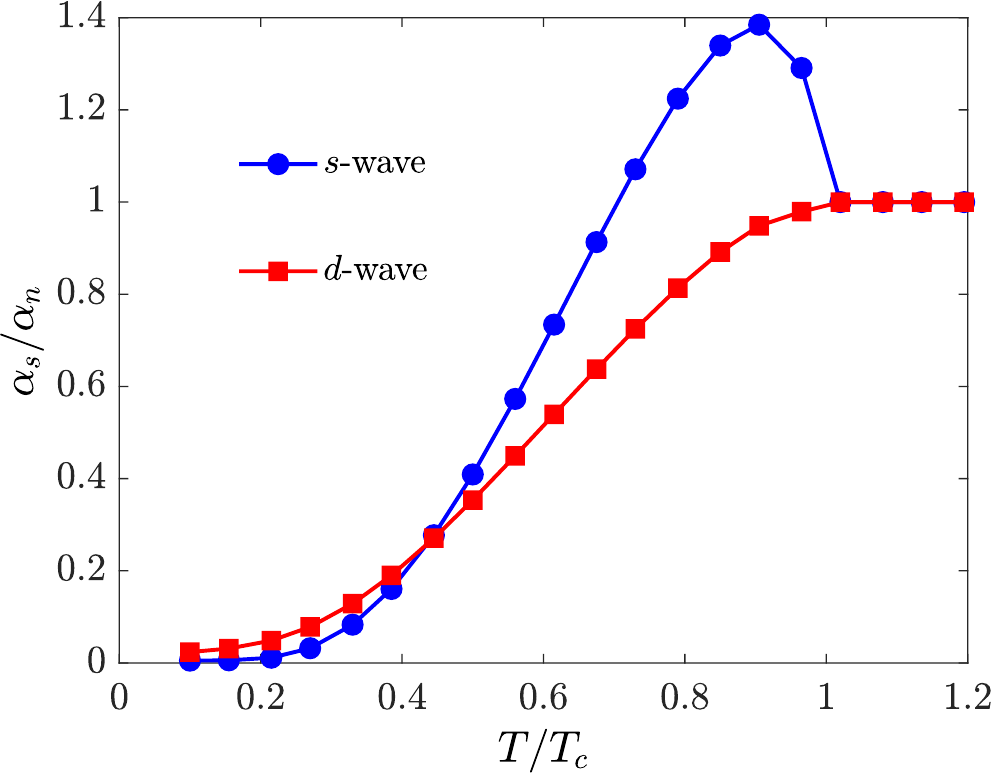}
    \caption{Temperature dependence of the spin-lattice relaxation rate ratio $\alpha_s/\alpha_n$ for the simple square lattice. Here, $\alpha_s$ and $\alpha_n$ denote the relaxation rates $1/T_1T$ of the superconducting state and normal states, respectively. All cases are plotted for $\mathcal{N} = 4\times 10^4$ and $\omega = \eta = 0.015$. As seen, the sign-changing gap of the $d_{x^2-y^2}$ form wipes out the Hebel-Slichter peak.}
    \label{sq_hebel}
\end{figure}
\begin{equation}
\label{eq2_21}
    \begin{split}
B(\mathbf{q},\mathbf{k}_n)=\frac{\Delta^*_{\mathbf{k}_n+\mathbf{q}}\Delta_{\mathbf{k}_n}}{E_{\mathbf{k}_n}E_{\mathbf{k}_n+\mathbf{q}}},
    \end{split}
\end{equation}
where $\mathbf{k}_n$ is a $\mathbf{k}$-point on the Fermi surface, here chosen by the point that gives the largest contribution to $\chi^{+-}_0(\mathbf{q},\omega)$. Figure~\ref{fig_sq_cancellation} shows that $\sum_{\mathbf{q}}B(\mathbf{q},\mathbf{k}_n)$ is a finite value between 0 and 1 in the $s$-wave case, which enhances $\chi^{+-}_0(\mathbf{q},\omega)$ and contributes to a Hebel-Slichter peak, as seen from Fig.~\ref{sq_hebel}. By contrast, for the $d$-wave case it gives no enhancement because the positive and negative contributions cancel, as evident from Fig.~\ref{fig_sq_cancellation}. Consequently, as is well-known, there is essentially no Hebel-Slichter peak for the $d$-wave order parameter as seen from  Fig.~\ref{sq_hebel}~\cite{Parker2007}.

\subsection{The kagome lattice}
Superconductivity on the kagome lattice can be described through the usual Nambu formalism
\begin{equation}
\label{eq_3_1}
    \begin{split}
    \mathcal{H} = \sum_{\mathbf{k}}\Psi^{\dagger}_{\mathbf{k} }\widehat{H} (\mathbf{k})\Psi_{\mathbf{k}},
    \end{split}
\end{equation}
where
\begin{equation}
\label{eq_3_2}
    \begin{split}
    \widehat{H}(\mathbf{k}) = \begin{pmatrix}
    H_0(\mathbf{k})&-\Delta(\mathbf{k})\\
    -\Delta(\mathbf{k})^{\dagger} & -H_0^{T}(-\mathbf{k})
    \end{pmatrix},
    \end{split}
\end{equation}
and $\Psi^\dagger_{\mathbf{k}} = \begin{pmatrix}c^{\dagger}_{\mathbf{k}\uparrow} & c_{-\mathbf{k}\downarrow}\end{pmatrix}$ with $c^{\dagger}_{\mathbf{k}\sigma} = \begin{pmatrix} c^{\dagger}_{\mathbf{k}\sigma A}&c^{\dagger}_{\mathbf{k}\sigma B}&c^{\dagger}_{\mathbf{k}\sigma C}\end{pmatrix}$.
For on-site (OS) Cooper pairing, the superconducting order parameters with $A_{1}$ ($s$-wave), and $E_2$ ($d_{x^2-y^2}$- and $d_{xy}$-wave) symmetries are~\cite{Sofie2023}
\begin{equation}
\label{eq_3_3}
    \begin{split}
    \Delta_\Gamma = \Delta_0 f_{\text{OS},\Gamma},
    \end{split}
\end{equation}
where
\begin{align}
\label{eq_3_4}
    f_{\text{OS},s} &= \frac{1}{\sqrt{3}}\begin{pmatrix}
        +1 & 0 & 0\\
        0 & +1 & 0\\
        0 & 0 & +1\\
    \end{pmatrix},\\
\label{eq_3_5}
    f_{\text{OS},d_{x^2-y^2}} &= \frac{1}{\sqrt{6}}\begin{pmatrix}
        +1 & 0 & 0\\
        0 & -2 & 0\\
        0 & 0 & +1\\
    \end{pmatrix},\\
\label{eq_3_6}
    f_{\text{OS},d_{xy}} &= \frac{1}{\sqrt{2}}\begin{pmatrix}
        +1 & 0 & 0\\
        0 & 0 & 0\\
        0 & 0 & -1\\
    \end{pmatrix}.
\end{align}
The $d_{x^2-y^2}$ and $d_{xy}$ orders belonging to the 2D $E_2$ irreducible representation can also be obtained via NN pairing. These harmonics, however, are zero on the Fermi surface near the upper van Hove point~\cite{RomerEA22,Sofie2023,Jinhui2024}, and since only states close to the Fermi surface contribute to the imaginary part of the spin susceptibility $\mathop{\text{Im}}\chi^{+-}_0({\mathbf{q}},\omega)$, NN pairing terms lead to small contributions and are not important for the results presented here. The main feature in the kagome lattice due to the sublattice interference in the normal state band structure is bound to the symmetry of the order parameter. Since all higher harmonics of the order parameter belong to the same irrep, these are also subject to the same effects. Hence,
we only show the on-site $d$-wave pairing terms. Additionally, below we will also address the time-reversal symmetry breaking superposition $d+id$ defined by
\begin{equation}
    \Delta_{d+id} = \frac{\Delta_0}{\sqrt{2}} \left(f_{\text{OS},d_{x^2-y^2}} + if_{\text{OS},d_{xy}} \right),\label{eq_3_7}
\end{equation}
which is expected to be the preferred solution at sufficiently low temperatures $T$ within the $E_2$ irreducible representation. In the following, we set $\Delta_0 = 0.2$. 

We can apply the unitary transformation that diagonalizes $H_0$ to transform the order parameter from sublattice space to band space
\begin{equation}
\label{eq_3_8}
    \begin{split}
    \Delta_{nm}(\mathbf{k}) = u^{\ast}_{n\alpha}(\mathbf{k})\Delta_{\alpha \beta} u^{\ast}_{m\beta}(-\mathbf{k}),
    \end{split}
\end{equation}
where $u_{n\alpha}(\mathbf{k})$ is the eigenstate of $H_0(\mathbf{k})$ in band $n$. Since only the middle band ($n=m=2$) crosses the Fermi surface near the upper van Hove point at $\mu=0$, we can ignore interband pairing and use an effective Hamiltonian with solely this band
\begin{equation}
\label{eq_3_9}
    \begin{split}
    {H}_{\text{eff}}(\mathbf{k}) = 
    \begin{pmatrix}
        \xi_2(\textbf{k})&-\Delta_{22}(\textbf{k})\\
        -\Delta^*_{22}(\textbf{k}) &-\xi_2(\textbf{k})\\
    \end{pmatrix}.
    \end{split}
\end{equation}
In Appendix \ref{ap:threeband}, we show that this approximation agrees well with the numerical result obtained from the full Hamiltonian $ \widehat{H}(\mathbf{k})$.
For simplicity, we use the notation $\xi_\mathbf{k}\equiv \xi_2(\mathbf{k})$ and $\Delta_{\mathbf{k}} \equiv \Delta_{22}(\mathbf{k})$ in the following expressions. The spin susceptibility of the kagome lattice is

\begin{align}
\label{eq_3_95}
    \chi^{+-}_0\!(\mathbf{q},\tau)=
    \frac{1}{\mathcal{N}^2}\sum_{\substack {\mathbf{k}\mathbf{k}'\\\alpha \beta}}\langle T_\tau c^{\dagger}_{\alpha,\mathbf{k} +  \mathbf{q}\uparrow}c_{\alpha,\mathbf{k}\downarrow}c^{\dagger}_{\beta,\mathbf{k}'  -  \mathbf{q}\downarrow}(\tau)c_{\beta,\mathbf{k}'\uparrow}(\tau)\rangle.
\end{align}
In order to simplify this expression, we can perform the transformation from sublattice to band space. In band space, the approximation that only the middle band is chosen can further simplify it. Additionally, by performing a Fourier transformation, the spin susceptibility can be written in the frequency domain 
\begin{widetext}
\begin{align}
   & \chi^{+-}_0(\mathbf{q},\omega) =
    \frac{1}{\mathcal{N}} \!  \sum_{\mathbf{k},E>0}   \! \left[ \left(1 \! - \! \frac{\xi_{\mathbf{k}}\xi_{\mathbf{k} \! + \! \mathbf{q}} \! + \! \Delta^*_{\mathbf{k} \! + \! \mathbf{q}}\Delta_{\mathbf{k}}}{E_{\mathbf{k}}E_{\mathbf{k} \! + \! \mathbf{q}}}\right)  \! \frac{1 \! - \! f(E_{\mathbf{k}}) \! - \! f(E_{\mathbf{k} \! + \! \mathbf{q}})}{\omega \! + \! E_{\mathbf{k} \! + \! \mathbf{q}} \! + \! E_{\mathbf{k}} \! + \! i\eta}\right.
     \! + \!  \left(1 \! - \! \frac{\xi_{\mathbf{k}}\xi_{\mathbf{k} \! + \! \mathbf{q}} \! + \! \Delta^*_{\mathbf{k} \! + \! \mathbf{q}}\Delta_{\mathbf{k}}}{E_{\mathbf{k}}E_{\mathbf{k} \! + \! \mathbf{q}}}\right) \! \frac{f(E_{\mathbf{k}}) \! + \! f(E_{\mathbf{k} \! + \! \mathbf{q}}) \! - \! 1}{\omega \! - \! E_{\mathbf{k} \! + \! \mathbf{q}} \! - \! E_{\mathbf{k}} \! + \! i\eta}
     \notag\\&
     \quad\quad\quad + \!   \left(1\! + \!\frac{\xi_{\mathbf{k}}\xi_{\mathbf{k} \! + \! \mathbf{q}}\! + \!\Delta^*_{\mathbf{k} \! + \! \mathbf{q}}\Delta_{\mathbf{k}}}{E_{\mathbf{k}}E_{\mathbf{k} \! + \! \mathbf{q}}}\right)\frac{f(E_{\mathbf{k}})\! - \!f(E_{\mathbf{k} \! + \! \mathbf{q}})}{\omega\! + \!E_{\mathbf{k} \! + \! \mathbf{q}}\! - \!E_{\mathbf{k}}\! + \!i\eta}
    \! + \!  \left. \left(1\! + \!\frac{\xi_{\mathbf{k}}\xi_{\mathbf{k} \! + \! \mathbf{q}}\! + \!\Delta^*_{\mathbf{k} \! + \! \mathbf{q}}\Delta_{\mathbf{k}}}{E_{\mathbf{k}}E_{\mathbf{k} \! + \! \mathbf{q}}}\right)\frac{f(E_{\mathbf{k} \! + \! \mathbf{q}})\! - \!f(E_{k})}{\omega\! + \!E_{\mathbf{k}}\! - \!E_{\mathbf{k} \! + \! \mathbf{q}}\! + \!i\eta}\right] \sum_{\alpha \beta} g_{\alpha\beta}(\mathbf{k},\mathbf{q}) ,
    \label{eq_3_10}
\end{align}
\end{widetext}
which is the same expression as for the square lattice, except for an extra factor $\sum_{\alpha,\beta}g_{\alpha\beta}(\mathbf{k},\mathbf{q})$ arising from the transformation from sublattice to band space with
\begin{equation}
\label{eq_3_11}
    \begin{split}
    g_{\alpha\beta}(\mathbf{k},\mathbf{q}) = u_{2\alpha}(\mathbf{k}+\mathbf{q})u_{2\beta }(\mathbf{k}+\mathbf{q})u_{2\alpha}(\mathbf{k})u_{2\beta}(\mathbf{k}).
    \end{split}
\end{equation}
We can write the eigenvectors without considering complex conjugation because they are real under the basis choice in Eq.~(\ref{eq1_2}). For the kagome lattice, it is customary to use another basis that is periodic in the first BZ. As discussed in Appendix \ref{ap:phasefactors}, care must be exerted for obtaining the spin susceptibility in that basis.

Similar to the discussion of the square lattice, we can define a (dressed) spin-susceptibility coherence factor given by 
\begin{equation}
\label{eq3_12}
    \begin{split}
    B_{\text{d}}(\mathbf{q},\mathbf{k}_n)=\frac{\sum_{\alpha \beta}g_{\alpha\beta}(\mathbf{k}_n,\mathbf{q})}{Z}\frac{\Delta^*_{\mathbf{k}_n+\mathbf{q}}\Delta_{\mathbf{k}_n}}{E_{\mathbf{k}_n}E_{\mathbf{k}_n+\mathbf{q}}},
    \end{split}
\end{equation}
where $Z$ is a normalization factor defined by
\begin{equation}
\label{eq3_13}
    \begin{split}
    Z=\frac{1}{\mathcal{N}^2}\sum_{\mathbf{k},\mathbf{q}} \sum_{\alpha \beta}g_{\alpha\beta}(\mathbf{k},\mathbf{q}).
    \end{split}
\end{equation}
For comparison, we also define a bare coherence factor without the matrix element dressing from $g_{\alpha\beta}(\mathbf{k},\mathbf{q})$
\begin{equation}
\label{eq3_14}
    \begin{split}
    B_{\text{b}}(\mathbf{q},\mathbf{k}_n)=\frac{\Delta^*_{\mathbf{k}_n+\mathbf{q}}\Delta_{\mathbf{k}_n}}{E_{\mathbf{k}_n}E_{\mathbf{k}_n+\mathbf{q}}}.
    \end{split}
\end{equation}
\begin{figure}[tb]
    \includegraphics[width=\linewidth]{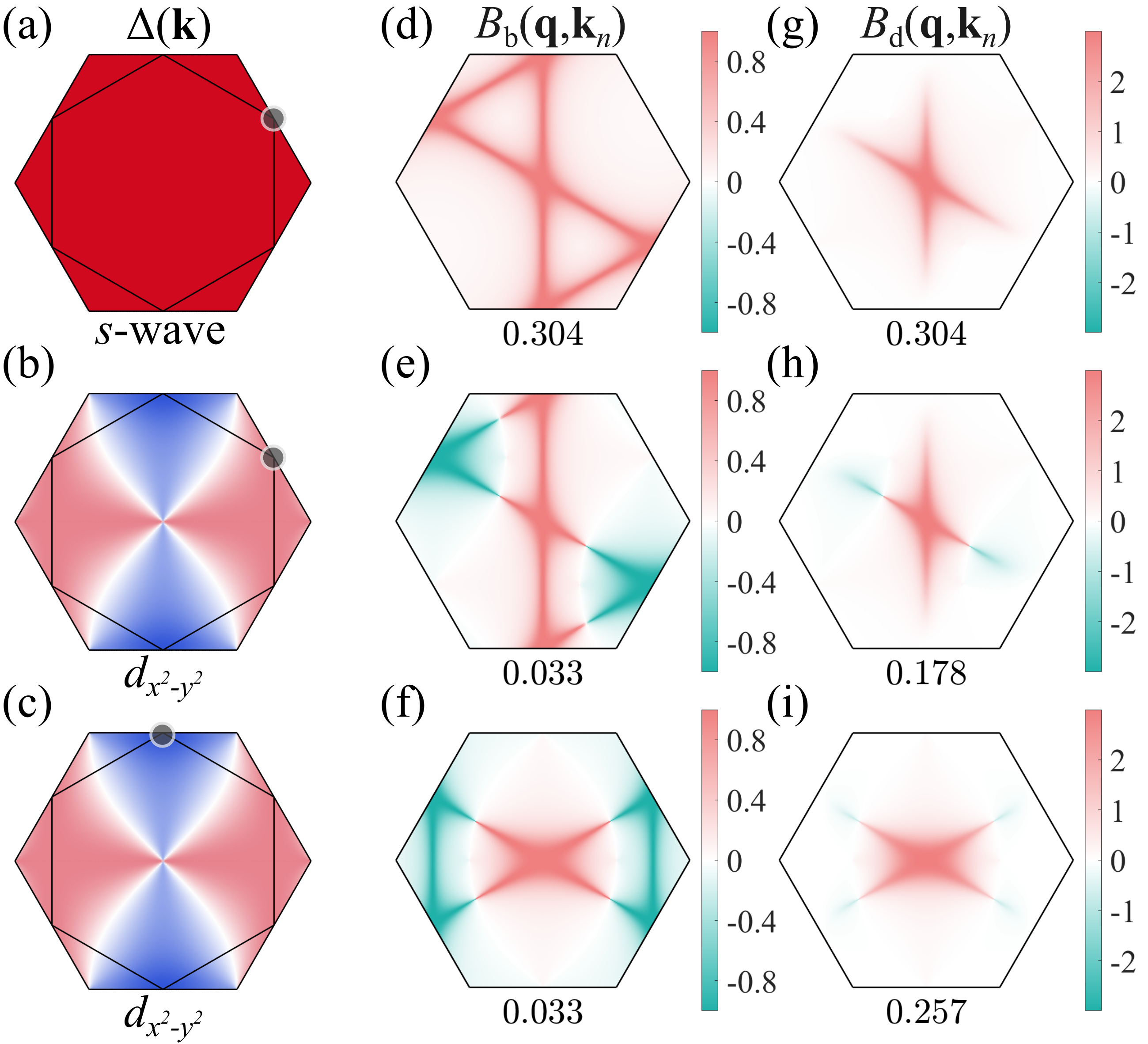}
    \caption{The $s$-wave order parameter (a) and the $d_{x^2-y^2}$ order parameter (b)-(c) in $\mathbf{k}$-space in the first BZ for the kagome lattice. The black lines indicate the Fermi surface at $\mu = 0$. Panel (d) displays the bare $B_{\text{b}}(\mathbf{q},\mathbf{k}_n)$ [Eq.~(\ref{eq3_14})]  for the $s$-wave order parameter in $\mathbf{q}$-space and panels (e)-(f) are for the $d_{x^2-y^2}$ order parameter. Panel (g) displays the dressed $B_{\text{d}}(\mathbf{q},\mathbf{k}_n)$ [Eq.~(\ref{eq3_12})] in $\mathbf{q}$-space for the $s$-wave order parameters and panels (h)-(i) are for the $d_{x^2-y^2}$ order parameter. The chosen $\mathbf{k}_n$ for (d)-(i) is indicated by the black dots in (a)-(c) in the same row. The numbers below (d)-(f) and (g)-(i) display the summed values $\sum_{\mathbf{q}}B_{\mathrm b}(\mathbf{q},\mathbf{k}_n)$ and $\sum_{\mathbf{q}}B_{\mathrm d}(\mathbf{q},\mathbf{k}_n)$, respectively.}
    \label{fig_kagome_cancellation}
\end{figure}
In Fig.~\ref{fig_kagome_cancellation} we show both the bare and dressed coherence factors for the $s$- and $d$-wave pairing states on the kagome lattice. Considering first the $A_{1}$ s-wave case seen in Fig.~\ref{fig_kagome_cancellation}(a,d,g), even though the dressing factor $g_{\alpha\beta}(\mathbf{k},\mathbf{q})$ changes the distribution of $B_{\mathrm b}(\mathbf{q},\mathbf{k}_n)$ in $\mathbf{q}$-space, the summed value $\sum_{\mathbf{q}}B_{\mathrm d}(\mathbf{q},\mathbf{k}_n)$ remains substantial and unchanged.  Therefore, the corresponding $1/T_1 T$ curve remains the same in both the bare and dressed cases of $s$-wave superconductivity, both exhibiting a Hebel-Slichter peak as expected, see Fig.~\ref{fig_kagome_barevsdress}.

\begin{figure}[tb]
    \includegraphics[width=0.88\linewidth]{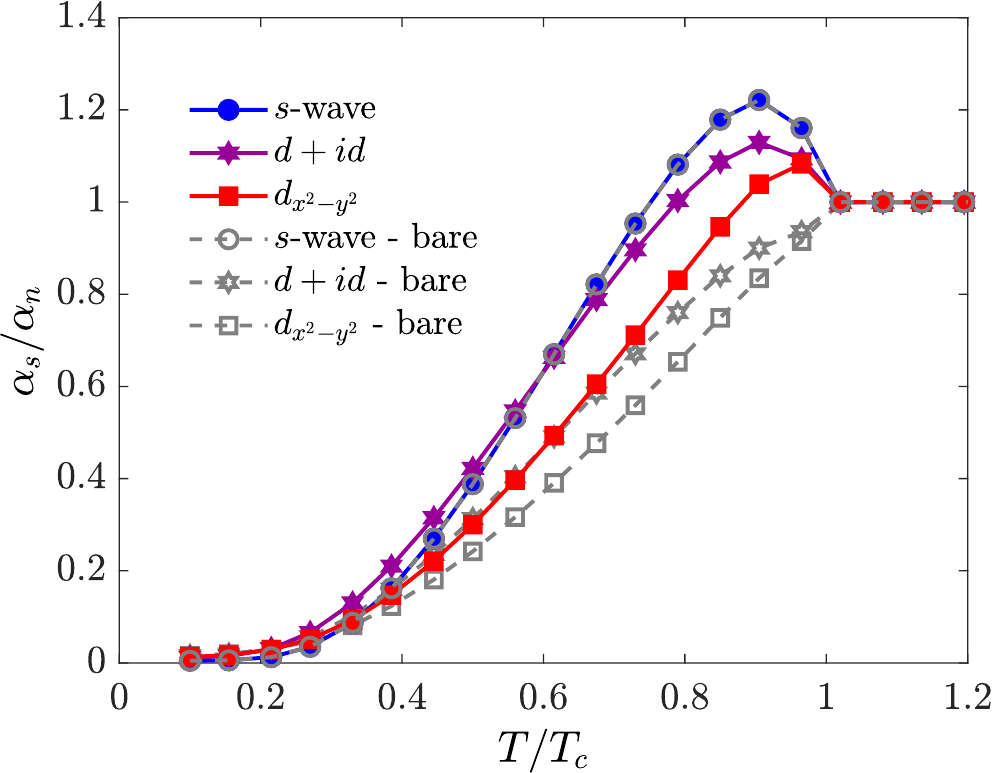}
    \caption{Temperature dependence of the spin-lattice relaxation rate ratio $\alpha_s/\alpha_n$ for superconductivity on the kagome lattice. All cases are plotted for $\mathcal{N} = 4\times 10^4$ and $\omega = \eta = 0.015$. The solid lines show the results calculated from the full $\chi^{+-}_0(\mathbf{q},\omega)$, while the dashed lines display the results obtained from the spin susceptibility without the dressing factor $g_{\alpha\beta}(\mathbf{k},\mathbf{q})$. As seen,  $g_{\alpha\beta}(\mathbf{k},\mathbf{q})$ restores the Hebel-Slichter peak even for the $d$-wave cases.}
    \label{fig_kagome_barevsdress}
\end{figure}
By contrast, for the $d$-wave case the dressing factor $g_{\alpha\beta}(\mathbf{k},\mathbf{q})$ originating from the sublattice to band space transformation becomes crucial, as seen from comparing Fig.~\ref{fig_kagome_cancellation}(e,f) to Fig.~\ref{fig_kagome_cancellation}(h,i). There we compare the coherence factors $B_{\mathrm b}(\mathbf{q},\mathbf{k}_n)$ and $B_{\mathrm d}(\mathbf{q},\mathbf{k}_n)$ for the two ${\mathbf{k}}$-points highlighted in Fig.~\ref{fig_kagome_cancellation}(b,c). Evidently, $g_{\alpha\beta}(\mathbf{k},\mathbf{q})$ destroys the compensation (the near cancellation between positive and negative regions) seen in  Fig.~\ref{fig_kagome_cancellation}(e,f) and leads to substantial summed values of $\sum_{\mathbf{q}}B_{\text{d}}(\mathbf{q},\mathbf{k}_n)$, as seen from Fig.~\ref{fig_kagome_cancellation}(h,i).
The origin of this effect is tied to the sublattice weights. More specifically, for the second band the sublattice weights vanish identically on extended regions of the BZ as seen from Fig.~\ref{fig1} (c). For the $d$-wave order parameters, the sign changes occur between regions of largest weight and regions with zero weight. Therefore, most of the sign change in the gap function is ``washed out'' by the product of the components of the eigenvectors in Eq.~(\ref{eq_3_11}). This conclusion is based on symmetry and therefore is not qualitatively altered if further neighbor pairing terms are present in the system, which we explicitly checked in separate calculations.
This implies that while the bare $d$-wave case should not exhibit a Hebel-Slichter peak, the full (dressed) case should. Indeed, as seen from  Fig.~\ref{fig_kagome_barevsdress} this is the case. As a result, there exists a pronounced Hebel-Slichter peak in the case of $d$-wave superconductivity on the kagome lattice. This is the main result of the present paper. This conclusion is both valid for the nodal $d$-wave cases and the nodeless TRSB $d+id$ order. For the $d+id$ case, $\sum_{\mathbf{q}}B_{\text{d}}(\mathbf{q},\mathbf{k}_n)$ remains real, even though the order parameter itself is complex. This is further discussed in Appendix \ref{ap:Bfactors}.

As seen from Fig.~\ref{fig_kagome_cancellation}, the summed values $\sum_{\mathbf{q}}B_{\mathrm d}(\mathbf{q},\mathbf{k}_n)$ are smaller in the dressed $d$-wave cases as compared to the $s$-wave case, resulting in slightly smaller Hebel-Slichter peaks in $d$-wave cases. This, however, is only a quantitative difference, with the important point being that, unlike the bare case or $d$-wave order on the square lattice, the Hebel-Slichter peak is not wiped out despite  the sign-changing gap. In summary, because of the kagome sublattice structure, which appears as the dressing factor $g_{\alpha\beta}(\mathbf{k},\mathbf{q})$, the enhancement factor $\sum_{\mathbf{q}}B_{\text{d}}(\mathbf{q},\mathbf{k}_n)$ in $\chi^{+-}_0(\mathbf{q},\omega)$ becomes substantial and $d$-wave superconductivity supports a Hebel-Slichter peak on the kagome lattice, as seen from Fig. \ref{fig_kagome_barevsdress}.

We note that the measurements of the NMR relaxation rate were done using Sb nuclear spins which are located in between the V atoms in the crystal structure. In this case the coupling between the nuclear spins and the conduction electrons is not a contact interaction and form factors appear in Eq.~(\ref{eq_2_2}). We derive these explicitly in Appendix \ref{ap:form_factor} and show that the qualitative conclusions are not altered by the additional form factor weighting of the momentum structure of the susceptibility.

\section{Discussion and conclusions}\label{sec:discussion}

We have demonstrated the existence of a Hebel-Slichter peak in the spin-lattice relaxation rate for $d$-wave superconductivity on the kagome lattice, and explained its existence from cancellation effects due to the peculiar sublattice-to-band space matrix elements.
This result remains qualitatively unchanged if the system is (i) slightly away from the the upper van Hove filling, or (ii) if further neighbor pairing terms are present. However, if band 1 forms the Fermi surface with the mixed-type sublattice structure, the sublattice weights do not vanish anywhere in the BZ, and the kagome system behaves as an ordinary superconductor. In this case, unconventional pairing states do not exhibit a Hebel-Slichter peak.
This result is in line with previous theoretical studies of the robustness of sign-changing gap structures to disorder on the kagome lattice~\cite{Sofie2023}.
Another example of unusual response is exemplified by the neutron resonance peak~\cite{ROSSATMIGNOD199186,Mook1993,Christianson2008,Inosov2010,StockNR}, i.e. a collective resonant spin state inside the $2\Delta$ gap of the spin susceptibility at some pronounced scattering vector ${\mathbf{q}}$, that depends on the band structure at hand. For the neutron resonance peak, it is the first two terms of Eqs.~\eqref{eq_2_1} and \eqref{eq_3_10} that are important. Therefore the relevant coherence factor is given by
\begin{equation}
    \left(1 \! - \! \frac{\xi_{\mathbf{k}}\xi_{\mathbf{k} \! + \! \mathbf{q}} \! + \! \Delta^*_{\mathbf{k} \! + \! \mathbf{q}}\Delta_{\mathbf{k}}}{E_{\mathbf{k}}E_{\mathbf{k} \! + \! \mathbf{q}}}\right),
\end{equation}
highlighting the absence (presence) of a neutron resonance mode for momentum vectors $\mathbf{q}$ connecting same-sign (opposite-sign) gap regions. For the kagome lattice, the factor $g_{\alpha\beta}(\mathbf{k},\mathbf{q})$ will significantly reduce contributions coming from $\Delta^*_{\mathbf{k} \! + \! \mathbf{q}}\Delta_{\mathbf{k}}$ even though $\mathbf{q}$ connect opposite signs of the gap. Therefore, the superconducting susceptibility is not expected to be significantly enhanced compared to the normal state, and in this sense the neutron resonance peak is wiped out. 

Similar unusual properties may be expected in other correlation functions for crystal structures where the important contributing states exhibit pronounced sublattice differentiation. For the kagome lattice, important effects of the matrix elements may be expected also for, for example, the behavior of the penetration depth and thermal conductivity since these quantities are obtained from two-particle correlation functions.

At present, it remains an open question what is the relevance of the findings in the present paper to the $A$V$_3$Sb$_5$ ($A$: K, Rb, Cs) and CsTi$_3$Bi$_5$ kagome superconductors under intense current investigations. To the best of our knowledge, these materials may turn out to host conventional $s$-wave superconductivity despite their many other unusual electronic properties~\cite{Wilsonreview}. However, the results in this paper, combined with Ref.~\cite{Sofie2023}, highlight the somewhat deceptive behavior of unconventional superconductivity on the kagome lattice, rendering it ``conventional'' in appearance. Specifically, $d$-wave superconductivity exhibits a Hebel-Slichter peak, no pronounced neutron resonance mode, and very weak $T_c$-suppression in response to nonmagnetic disorder. Therefore, from this perspective, the question of the pairing symmetry of the above-mentioned kagome superconductors remains open at present.

\begin{acknowledgments}
We acknowledge fruitful discussions with M. H. Christensen, P. J. Hirschfeld and S. C. Holb\ae k. A.K. acknowledges support by the Danish National Committee for Research Infrastructure (NUFI) through the ESS-Lighthouse Q-MAT.
\end{acknowledgments}

\appendix
\renewcommand{\thefigure}{S\arabic{figure}}
\setcounter{figure}{0}

\section{Three-band model without approximation}\label{ap:threeband}

\begin{figure}[tb]
    \centering
    \includegraphics[width=0.88\linewidth]{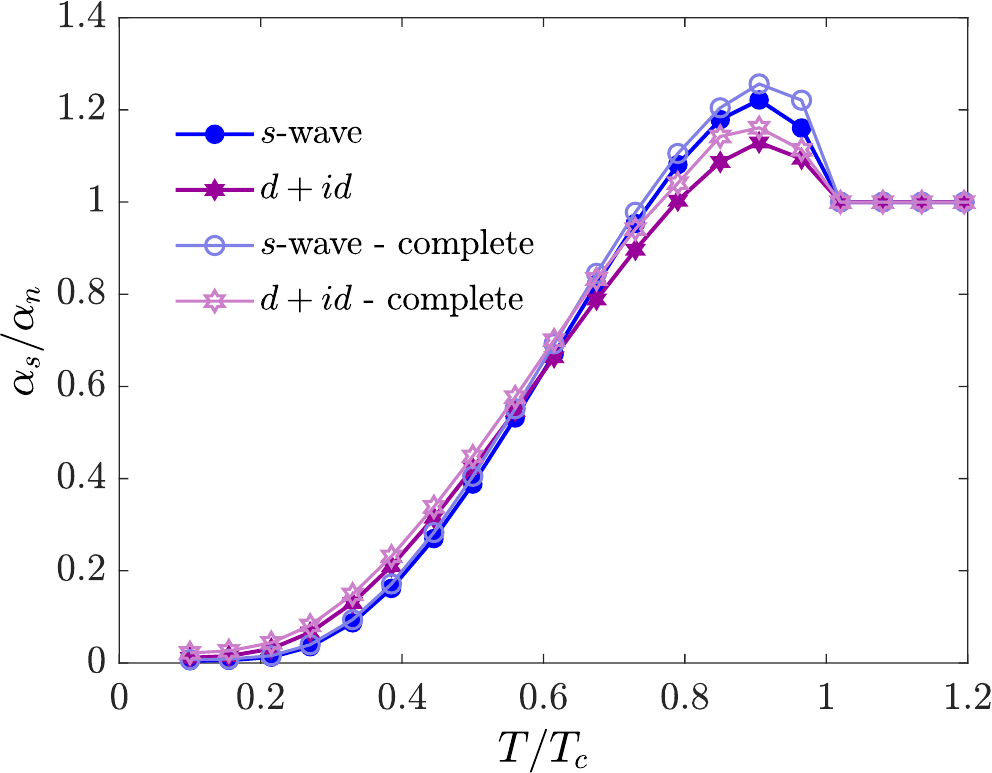}
    \caption{Comparison of calculation including all bands (``complete'', Eq.~(\ref{eq_chi_3_band})) with the single band approximation, Eq.~(\ref{eq_3_10}).
    The temperature dependence of the spin-lattice relaxation rate ratio $\alpha_s/\alpha_n$ for the kagome system does only exhibit minor qualitative differences from the single-band approximation. Calculation for the single band approximation is done for $\mathcal{N} = 4\times 10^4$ and the complete calculation for $\mathcal{N} = 1\times 10^4$, respectively; $\omega = \eta = 0.015$ is used in both cases. The blue and purple lines show the results calculated from the effective Hamiltonian $H_\text{eff}(\mathbf{k})$. The respective lines with open symbols show the results obtained from the complete three-band Hamiltonian $\widehat{H}(\mathbf{k})$.}
    \label{fig_appendix_c}
\end{figure}

Apart from performing the sublattice-to-band transformation shown in Eqs.~\eqref{eq_3_8} and \eqref{eq_3_9}, which results in the dressing factor $g_{\alpha\beta}(\mathbf{k},\mathbf{q})$, one can also numerically diagonalize the BdG Hamiltonian Eq.~\eqref{eq_3_2} in sublattice space directly. The matrix that diagonalizes $\widehat{H}(\mathbf{k})$ through $\mathbf{U}^{-1}(\mathbf{k})\widehat{H}(\mathbf{k})\mathbf{U}(\mathbf{k})$ is
\begin{widetext}
\begin{equation}
\label{eq_c1}
    \begin{split}
    &\mathbf{U}(\mathbf{k}) =
    \begin{pmatrix}
        u_{A,1,\mathbf{k}}^* & u_{A,2,\mathbf{k}}^* & u_{A,3,\mathbf{k}}^*&-v_{A,1,\mathbf{k}} & -v_{A,2,\mathbf{k}} & -v_{A,3,\mathbf{k}}\\
        u_{B,1,\mathbf{k}}^* & u_{B,2,\mathbf{k}}^* & u_{B,3,\mathbf{k}}^*&-v_{B,1,\mathbf{k}} & -v_{B,2,\mathbf{k}} & -v_{B,3,\mathbf{k}}\\
        u_{C,1,\mathbf{k}}^* & u_{C,2,\mathbf{k}}^* & u_{C,3,\mathbf{k}}^*&-v_{C,1,\mathbf{k}} & -v_{C,2,\mathbf{k}} & -v_{C,3,\mathbf{k}}\\
        v_{A,1,\mathbf{k}}^{*} & v_{A,2,\mathbf{k}}^{*} & v_{A,3,\mathbf{k}}^{*} & u_{A,1,\mathbf{k}} & u_{A,2,\mathbf{k}} & u_{A,3,\mathbf{k}}\\
        v_{B,1,\mathbf{k}}^{*} & v_{B,2,\mathbf{k}}^{*} & v_{B,3,\mathbf{k}}^{*} & u_{B,1,\mathbf{k}} & u_{B,2,\mathbf{k}} & u_{B,3,\mathbf{k}}\\
        v_{C,1,\mathbf{k}}^{*} & v_{C,2,\mathbf{k}}^{*} & v_{C,3,\mathbf{k}}^{*} & u_{C,1,\mathbf{k}} & u_{C,2,\mathbf{k}} & u_{C,3,\mathbf{k}}\\
    \end{pmatrix},
    \end{split}
\end{equation}
which can be obtained numerically. The spin susceptibility in this basis is
    \begin{align}
    \label{eq_chi_3_band}
    &\chi^{+-}_{0}(\textbf{q},\omega)
    =\frac{1}{\mathcal{N}}\!\sum_{\substack{\mathbf{k},m,m' \\\alpha \beta}}\!
    \left[
    (v^*_{\alpha,m,\mathbf{k}+\mathbf{q}} v_{\beta,m,\mathbf{k}+\mathbf{q}}u^*_{\alpha,m',\mathbf{k}}u_{\beta,m',\mathbf{k}}-v^*_{\alpha,m,\mathbf{k}+\mathbf{q}}u_{\beta,m,\mathbf{k}+\mathbf{q}}u^*_{\alpha,m',\mathbf{k}}v_{\beta,m',\mathbf{k}}) \frac{1-f(E_{\mathbf{k}+\mathbf{q},m})-f(E_{\mathbf{k},m'})}{\omega+E_{\mathbf{k}+\mathbf{q},m}+E_{\mathbf{k},m'}+i\eta}\right. \notag\\
    &+(u_{\alpha,m,\mathbf{k}+\mathbf{q}}u^*_{\beta,m,\mathbf{k}+\mathbf{q}}u^*_{\alpha,m',\mathbf{k}}u_{\beta,m',\mathbf{k}} +u_{\alpha,m,\mathbf{k}+\mathbf{q}}v^*_{\beta,m,\mathbf{k}+\mathbf{q}}u^*_{\alpha,m',\mathbf{k}}v_{\beta,m',\mathbf{k}}) \frac{f(E_{\mathbf{k}+\mathbf{q},m})-f(E_{\mathbf{k},m'})}{\omega+E_{\mathbf{k},m'}-E_{\mathbf{k}+\mathbf{q},m}+i\eta}\notag\\
    &+ (v^*_{\alpha,m,\mathbf{k}+\mathbf{q}} v_{\beta,m,\mathbf{k}+\mathbf{q}}v_{\alpha,m',\mathbf{k}}v^*_{\beta,m',\mathbf{k}}+v^*_{\alpha,m,\mathbf{k}+\mathbf{q}}u_{\beta,m,\mathbf{k}+\mathbf{q}}v_{\alpha,m',\mathbf{k}}u^*_{\beta,m',\mathbf{k}}) \frac{f(E_{\mathbf{k},m'})-f(E_{\mathbf{k}+\mathbf{q},m})}{\omega+E_{\mathbf{k}+\mathbf{q},m}-E_{\mathbf{k},m'}+i\eta}\notag\\
    &\left. +(u_{\alpha,m,\mathbf{k}+\mathbf{q}}u^*_{\beta,m,\mathbf{k}+\mathbf{q}}v_{\alpha,m',\mathbf{k}}v^*_{\beta,m',\mathbf{k}}-u_{\alpha,m,\mathbf{k}+\mathbf{q}}v^*_{\beta,m,\mathbf{k}+\mathbf{q}}v_{\alpha,m',\mathbf{k}}u^*_{\beta,m',\mathbf{k}}) \frac{f(E_{\mathbf{k}+\mathbf{q},m})+f(E_{\mathbf{k},m'})-1}{\omega-E_{\mathbf{k}+\mathbf{q},m}-E_{\mathbf{k},m'}+i\eta}\right].
\end{align}
\end{widetext}
There is no approximation in this expression. We compare the results obtained from two methods in Fig.~\ref{fig_appendix_c}. As seen, they are very similar despite the approximation made in Eq.~\eqref{eq_3_9} that only considers the middle band relevant at the Fermi level. The result from the complete model is slightly larger than the effective model because of small contributions of the other bands. Thus, the Hebel-Slichter peak is larger in the complete model compared to the effective model.

\section{Basis choice and associated phase factors}\label{ap:phasefactors}
The Hamiltonian in Eq.~\eqref{eq1_2} is not periodic in the first BZ, meaning that while the eigenvalues are periodic, the eigenvectors are not. We can remedy this issue by choosing the size of the $\mathbf{k}$-grid twice that of the first BZ. Alternatively, we can apply the unitary transformation $T^{-1}(\mathbf{k})H_{0}(\mathbf{k})T(\mathbf{k})$, where
\begin{equation}
    \label{eq_a1}
    \begin{split}
    T(\mathbf{k}) = 
    \begin{pmatrix}
        e^{-ik_1}&0&0\\
        0&e^{-ik_2}&0\\
        0&0&1
    \end{pmatrix},
    \end{split}
\end{equation}
and the Hamiltonian matrix in the new basis becomes 
\begin{align}
\label{eq_a2}
    &\Tilde{H}_{0}(\mathbf{k})
    =  -
    \begin{pmatrix}
        \mu&t(1\! + \!e^{2ik_3})&t(1\! + \!e^{-2ik_1})\\
        t(1\! + \!e^{-2ik_3})&\mu&t(1\! + \!e^{-2ik_2})\\
        t(1\! + \!e^{2ik_1})&t(1\! + \!e^{2ik_2})&\mu
    \end{pmatrix},
\end{align}
which is periodic in the first BZ. The Nambu Hamiltonian can also be transformed to the new basis by $T_{\text{Nambu}}^{-1}(\mathbf{k})\widehat{H}(\mathbf{k})T_{\text{Nambu}}(\mathbf{k})$, where
\begin{equation}
    \label{eq_a3}
    \begin{split}
    T_{\text{Nambu}}(\mathbf{k}) = 
    \begin{pmatrix}
        T(\mathbf{k}) & 0\\
        0 & T^{T}(\mathbf{-k})
    \end{pmatrix}.
    \end{split}
\end{equation}
This transformation introduces extra phase factors in the spin susceptibility. We can recover the susceptibility under basis Eq. \eqref{eq1_2} by considering the phase factors
\begin{align}
\label{eq_a4}
    \chi^{+-}_{0,\alpha\beta}(\mathbf{q},\omega)
    \!=\!
    \begin{pmatrix}
       \! \tilde{\chi}^{+-}_{0,AA} \!\!& \!\!e^{i(q_2-q_1)}\tilde{\chi}^{+-}_{0,AB} &e^{-iq_1}\tilde{\chi}^{+-}_{0,AC}\\
       \! e^{i(q_1-q_2)}\tilde{\chi}^{+-}_{0,BA} \!\!& \!\!\tilde{\chi}^{+-}_{0,BB} &e^{-iq_2}\tilde{\chi}^{+-}_{0,BC}\\
       \! e^{iq_1}\tilde{\chi}^{+-}_{0,CA}\!\! &\! \!e^{iq_2}\tilde{\chi}^{+-}_{0,CB} &\tilde{\chi}^{+-}_{0,CC}\\
    \end{pmatrix}\!,
\end{align}
where $\tilde{\chi}^{+-}_{0,\alpha\beta}(\mathbf{q},\omega)$ is the spin susceptibility in the new basis, and $q_n = \mathbf{q}\cdot \mathbf{a}_n$.

\begin{figure}[tb]
    \includegraphics[width=\linewidth]{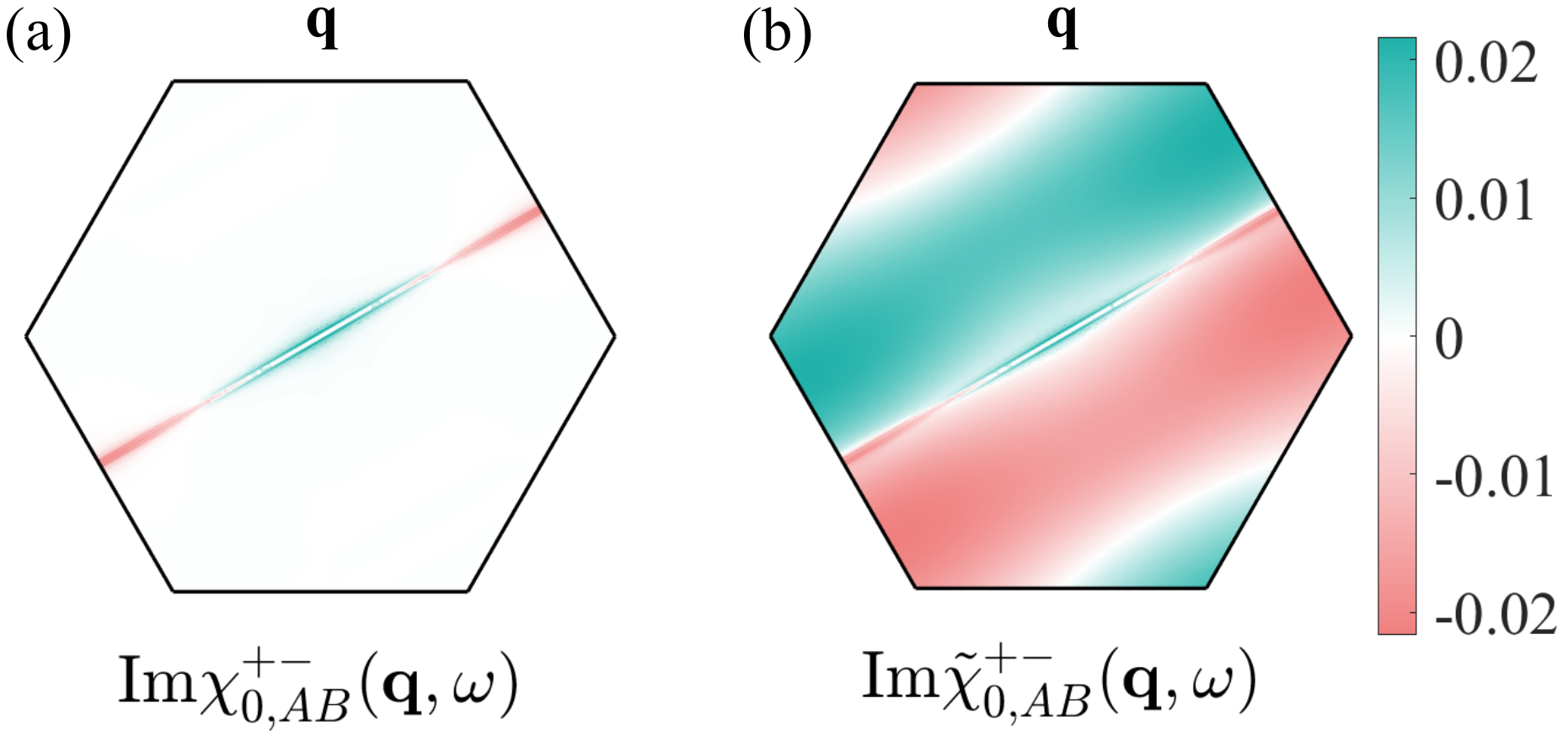}
    \caption{Panel (a) displays $\mathop{\text{Im}}{\chi}^{+-}_{0,AB}(\mathbf{q},\omega)$ with the basis Eq.~(\ref{eq1_2}). Panel (b) displays $\mathop{\text{Im}}\tilde{\chi}^{+-}_{0,AB}(\mathbf{q},\omega)$ with the basis choice of Eq.~(\ref{eq_a2}). (a) conserves the $C_2$ symmetry by a rotation of $\pi$ around the principal axis; (b) breaks this symmetry.}
    \label{AppendixA}
\end{figure}

It is straightforward to use the basis in Eq.~\eqref{eq1_2} for spin susceptibility calculations, while direct use of the basis Eq.~\eqref{eq_a2} breaks, for example, the $C_2$ symmetry by a rotation of $\pi$ around the principal axis. As shown in Fig.~\ref{AppendixA}(a), the spin susceptibility for sublattice index $AB$ calculated from the basis in Eq.~\eqref{eq1_2} is invariant under a $C_2$ rotation, which is a symmetry of the Hamiltonian. Figure~\ref{AppendixA}(b) shows that the basis Eq.~\eqref{eq_a2} breaks this symmetry because the negative and positive parts cannot be mapped onto each other by this symmetry operation.  However, by including the phase factors of Eq.~\eqref{eq_a4} one recovers the correct susceptibility. This symmetry breaking by an inappropriate choice of basis is also discussed in Ref.~\cite{Sumita2023}.

\section{Absence of imaginary part of $B$ factor}\label{ap:Bfactors}

For $d+id$ superconductivity, $\Delta_{d+id}$ is a complex number. Thus, in principle, the imaginary part of $B(\mathbf{q},\mathbf{k})$ might contribute to $\chi^{+-}_0(\mathbf{q},\omega)$. If only the real part of $B(\mathbf{q},\mathbf{k})$ contributes to $\chi^{+-}_0(\mathbf{q},\omega)$, then only the imaginary part of the Fermi function term, which is only non-zero near the Fermi surface, contributes to the imaginary part of $\chi^{+-}_0(\mathbf{q},\omega)$. We define 
\begin{equation}
\label{eq_b1}
    \begin{split}
    \Delta_{\mathbf{k}} = \Delta^{\prime}_\mathbf{k} + i  \Delta^{\prime\prime}_\mathbf{k},
    \end{split}
\end{equation}
where $\Delta^{\prime}_{\mathbf{k}}$ and $\  \Delta^{\prime\prime}_\mathbf{k}$ denote the real and  imaginary parts of $\Delta_\mathbf{k}$. Both $\Delta^{\prime}_\mathbf{k}$ and $\  \Delta^{\prime\prime}_\mathbf{k}$ have the time-reversal symmetry, while $\Delta_{\mathbf{k}}$ breaks the time-reversal symmetry.  $B(\mathbf{q},\mathbf{k})$ can be written as
\begin{equation}
\label{eq_b2}
    \begin{split}
    B(\mathbf{q},\mathbf{k}) &= \frac{(\Delta^{\prime}_{\mathbf{k}+\mathbf{q}} - i  \Delta^{\prime\prime}_{\mathbf{k}+\mathbf{q}}) (\Delta^{\prime}_{\mathbf{k}} + i  \Delta^{\prime\prime}_{\mathbf{k}})}{E_{\mathbf{k}}E_{\mathbf{k}+\mathbf{q}}},
    \end{split}
\end{equation}
and the imaginary part is 
\begin{align}
\label{eq_b3}
    \mathop{\text{Im}}B(\mathbf{q},\mathbf{k})= \frac{\Delta^{\prime}_{\mathbf{k}+\mathbf{q}}\Delta^{\prime\prime}_{\mathbf{k}} - \Delta^{\prime\prime}_{\mathbf{k}+\mathbf{q}}\Delta^{\prime}_{\mathbf{k}}}{E_{\mathbf{k}}E_{\mathbf{k}+\mathbf{q}}}.
\end{align}
Since we sum over $\mathbf{k}$ in $\chi^{+-}_0(\mathbf{q},\omega)$, we can do $\mathbf{k} \rightarrow -\mathbf{k} - \mathbf{q}$ for the first term of Eq. \eqref{eq_b3}. Due to the even parity property $\Delta^{\prime}_{\mathbf{k}}=\Delta^{\prime}_{-\mathbf{k}}$, $\Delta^{\prime\prime}_{\mathbf{k}} = \Delta^{\prime\prime}_{-\mathbf{k}}$ and $E_{\mathbf{k}}=E_{-\mathbf{k}}$ together, the imaginary part gives zero. Thus for $\Delta_{d+id}$, we can still only consider $\mathop{\text{Re}}B(\mathbf{q},\mathbf{k})$.

\section{Form factors}\label{ap:form_factor}

\begin{figure}[tb]
    \includegraphics[width=0.5\linewidth]{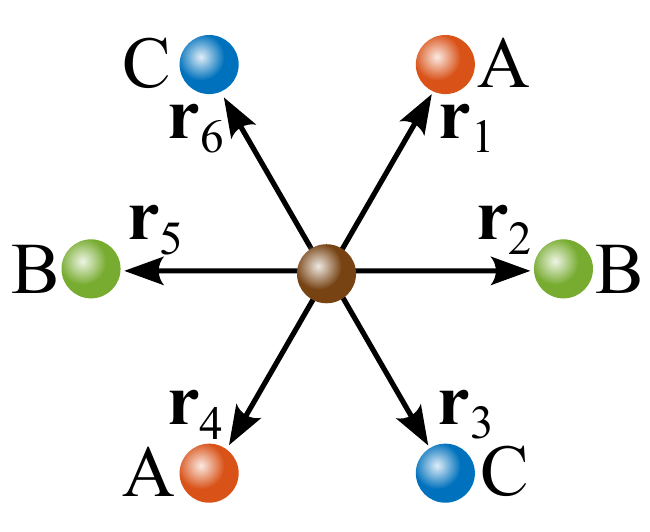}
    \caption{Illustration of the in-plane antimony atom in the kagome superconductors $A$V$_3$Sb$_5$. The brown site refers to the antimony atom, and the colored ones refer to the vanadium atoms on three different sublattice sites.}
    \label{antimony}
\end{figure}

The general expression for the NMR relaxation rate is derived from  perturbation theory \cite{Moriya1963Effect,Smerald2011NMR}. The perturbation is described by the hyperfine coupling interaction
\begin{equation}
\label{eq_d1}
    \begin{split}
    \mathcal{H}_{\text{hf}} = \sum_{i} \mathbf{I}\cdot \overline{\overline{\mathcal{A}}}_{i} \cdot \mathbf{S}_{i\alpha},
    \end{split}
\end{equation}
where $\mathbf{I}$ and $\mathbf{S}_{i\alpha}$ denote the nuclear and electron spins. The index $i$ labels the sites to which the nuclear spin couples. In the case of V NMR, it is sufficient to consider only the onsite term, while for Sb NMR, the sum runs over the nearest neighbor V atoms where conducting electrons couple to the nuclear spin as shown in Fig.~\ref{antimony}. The index $\alpha$ labels the corresponding sublattice indices on those sites. $\overline{\overline{\mathcal{A}}}_{i}$ is the hyperfine coupling matrix describing
the coupling between the nuclear and electron spins, which can be written in Cartesian coordinates as
\begin{equation}
\label{eq_d2}
    \begin{split}
    \overline{\overline{\mathcal{A}}}_{i} = 
    \begin{pmatrix}
        \mathcal{A}^{xx}_{i}&\mathcal{A}^{xy}_{i}&\mathcal{A}^{xz}_{i}\\
        \mathcal{A}^{yx}_{i}&\mathcal{A}^{yy}_{i}&\mathcal{A}^{yz}_{i}\\
        \mathcal{A}^{zx}_{i}&\mathcal{A}^{zy}_{i}&\mathcal{A}^{zz}_{i}
    \end{pmatrix}.
    \end{split}
\end{equation}
In principle, the elements of this matrix can be measured by Knight-shift experiments, but components of the tensor for different $i$ are related by the crystal symmetries which we use in the following. To derive the effect of the elements of the coupling matrix, we start from the general expression for spin-lattice relaxation rate as
\begin{align}
\label{eq_d3}
    \frac{1}{T_1} 
    \propto \int_{-\infty}^{\infty} dt\  \cos(\omega_0 t)\left(\langle  \{h_{x}(t) , h_{x}\}  + \{h_{y}(t) , h_{y}\} \rangle\right) ,
\end{align}
where $h_{x}$ and $h_{y}$ are the $x$ and $y$ components of 
\begin{equation}
\label{eq_d4}
    \begin{split}
    \mathbf{h} = \sum_{i} \overline{\overline{\mathcal{A}}}_{i} \cdot \mathbf{S}_{i\alpha}.
    \end{split}
\end{equation}
By performing a Fourier transformation for the electron spins
\begin{equation}
\label{eq_d5}
    \begin{split}
    \mathbf{S}_{i\alpha} = \frac{1}{\sqrt{\mathcal{N}}}\sum_{\mathbf{q}}e^{i\mathbf{q}\mathbf{r}_i}\mathbf{S}_{\mathbf{q} \alpha},
    \end{split}
\end{equation}
the relaxation rate can be written as
\begin{equation}
\label{eq_d6}
    \begin{split}
    &\frac{1}{T_1} 
    \propto \int_{-\infty}^{\infty} dt\  \frac{1}{\mathcal{N}} \sum_{\mathbf{q}} \cos(\omega_0 t)\\\times&\left(\left\langle  \Bigl\{\bigl[\sum_{i} \overline{\overline{\mathcal{A}}}_{i} \cdot e^{i\mathbf{q}\cdot\mathbf{r}_i} \mathbf{S}_{\mathbf{q}\alpha}\bigl]^{x}(t) ,\bigl[\sum_{j} \overline{\overline{\mathcal{A}}}_{j} \cdot e^{-i\mathbf{q}\cdot\mathbf{r}_j} \mathbf{S}_{-\mathbf{q}\beta}\bigl]^{x}\Bigl\}  \right.\right. \\
    + &\left. \left. \Bigl\{\bigl[\sum_{i} \overline{\overline{\mathcal{A}}}_{i} \cdot e^{i\mathbf{q}\cdot\mathbf{r}_i} \mathbf{S}_{\mathbf{q}\alpha}\bigl]^{y}(t) ,\bigl[\sum_{j} \overline{\overline{\mathcal{A}}}_{j} \cdot e^{-i\mathbf{q}\cdot\mathbf{r}_j} \mathbf{S}_{-\mathbf{q}\beta}\bigl]^{y}\Bigl\} \right\rangle\right) .
    \end{split}
\end{equation}

If the conducting electrons couple to the nuclear spin on the same site, such as the vanadium atoms in $A$V$_3$Sb$_5$ kagome metals, this expression will reduce to Eq.~\eqref{eq_2_2}. However, if the conducting electrons
are localized on different sites to the nuclear spins, for example the antimony atoms in $A$V$_3$Sb$_5$, an additional form factor will enter the expression because of the exponential coefficient in the Fourier transformation \cite{Shastry1989form,Mila1989form,Smerald2011NMR}. In fact, in the experiment which measures the spin-lattice relaxation rate of CsV$_3$Sb$_5$ in the superconducting state, the measured ones are the antimony atoms which reside on different sites to the active electron degrees of freedom \cite{Mu2021S-wave}. Thus to directly connect to the experimental setup, an additional form factor should be considered.

To relate coupling matrices to different nearest neighbors, we use that the Hamiltonian of the kagome lattice is invariant under $C_6$ rotations in the $xy$ plane. The 60$^{\circ}$ clockwise rotation $\overline{\overline{\mathcal{M}}}$ in the $xy$ plane can be expressed as
\begin{equation}
\label{eq_d7}
    \begin{split}
    \overline{\overline{\mathcal{M}}} = 
    \begin{pmatrix}
        \frac{1}{2}&\frac{\sqrt{3}}{2}&0\\
        -\frac{\sqrt{3}}{2}&\frac{1}{2}&0\\
        0&0&1
    \end{pmatrix}.
    \end{split}
\end{equation}
For simplicity, we assume that the first coupling matrix is a constant $1$ in all directions
\begin{equation}
\label{eq_d8}
    \begin{split}
    \overline{\overline{\mathcal{A}}}_{1} = 
    \begin{pmatrix}
        1&1&1\\
        1&1&1\\
        1&1&1\\
    \end{pmatrix}.
    \end{split}
\end{equation}
To keep the hyperfine Hamiltonian invariant under $C_6$ rotation, the second coupling matrix should be
\begin{equation}
\label{eq_d9}
    \begin{split}
    \overline{\overline{\mathcal{A}}}_{2} = \overline{\overline{\mathcal{M}}} \overline{\overline{\mathcal{A}}}_{1} \overline{\overline{\mathcal{M}}}^{-1}= 
    \begin{pmatrix}
        \frac{2+\sqrt{3}}{2}&-\frac{1}{2}&\frac{1+\sqrt{3}}{2}\\
        -\frac{1}{2}&\frac{2-\sqrt{3}}{2}&\frac{1-\sqrt{3}}{2}\\
        \frac{1+\sqrt{3}}{2}&\frac{1-\sqrt{3}}{2}&1
    \end{pmatrix}.
    \end{split}
\end{equation}
Similarly, the other coupling matrices can be derived

\begin{align}
\label{eq_d10}
    \overline{\overline{\mathcal{A}}}_{3} &= 
    \begin{pmatrix}
        \frac{2-\sqrt{3}}{2}&-\frac{1}{2}&\frac{\sqrt{3}-1}{2}\\
        -\frac{1}{2}&\frac{2+\sqrt{3}}{2}&\frac{-1-\sqrt{3}}{2}\\
        \frac{\sqrt{3}-1}{2}&\frac{-1-\sqrt{3}}{2}&1
    \end{pmatrix},\\
\label{eq_d11}
    \overline{\overline{\mathcal{A}}}_{4} &= 
    \begin{pmatrix}
        1&1&-1\\
        1&1&-1\\
        -1&-1&1
    \end{pmatrix},\\
\label{eq_d12}
    \overline{\overline{\mathcal{A}}}_{5} &= 
    \begin{pmatrix}
        \frac{2+\sqrt{3}}{2}&-\frac{1}{2}&\frac{-\sqrt{3}-1}{2}\\
        -\frac{1}{2}&\frac{2-\sqrt{3}}{2}&\frac{\sqrt{3}-1}{2}\\
        \frac{-\sqrt{3}-1}{2}&\frac{\sqrt{3}-1}{2}&1
    \end{pmatrix},\\
\label{eq_d13}
    \overline{\overline{\mathcal{A}}}_{6} &= 
    \begin{pmatrix}
        \frac{2-\sqrt{3}}{2}&-\frac{1}{2}&\frac{1-\sqrt{3}}{2}\\
        -\frac{1}{2}&\frac{2+\sqrt{3}}{2}&\frac{1+\sqrt{3}}{2}\\
        \frac{1-\sqrt{3}}{2}&\frac{1+\sqrt{3}}{2}&1
    \end{pmatrix}.
\end{align}

\begin{figure}[t]
    \centering
    \includegraphics[width=0.88\linewidth]{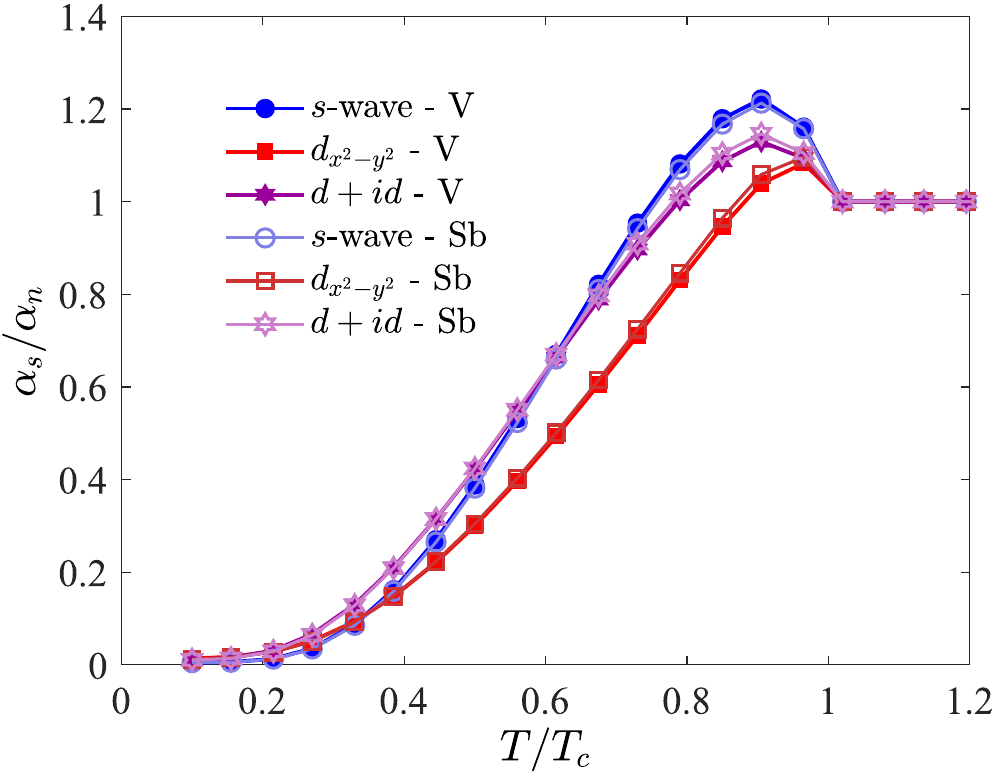}
    \caption{Comparison of expected spin relaxation rate in a NMR experiment on V nuclear spins and Sb nuclear spins. The temperature dependence of the spin-lattice relaxation rate ratio $\alpha_s/\alpha_n$ exhibits only tiny quantitative differences from the momentum dependent form factors in Eq.~(\ref{eq_d14}). All calculations are done with $\mathcal{N} = 4\times 10^4$ and $\omega = \eta = 0.015$ and using the  Hamiltonian $H_\text{eff}(\mathbf{k})$.
    The solid lines with full symbols correspond to the relaxation rate for the vanadium atoms. The respective lines with open symbols show the results including the form factors, which corresponds to the relaxation rate for the antimony atoms.}
    \label{fig_appendix_D}
\end{figure}

Inserting those coupling matrices into Eq.~\eqref{eq_d6}, we can get the expression of the relaxation rate for antimony atoms
\begin{equation}
\label{eq_d14}
    \begin{split}
    \alpha \propto \lim_{\omega\rightarrow 0} \frac{1}{\mathcal{N}}\sum_{\substack {\mathbf{q}\\\alpha \beta}} \mathcal{F}_{\alpha\beta}(\mathbf{q})\mathop{\text{Im}}\frac{\chi^{+-}_{0,\alpha\beta}(\mathbf{q},\omega)}{\omega},\\
    \end{split}
\end{equation}
where the momentum-dependent form factors are given by
\begin{align}
\label{eq_d15}
    \mathcal{F}_{AA} & = 16\cos^2 \!\left(\frac{1}{4}q_x\!+\!\frac{\sqrt{3}}{4}q_y\right) \!+\! 8 \sin^2 \!\left(\frac{1}{4}q_x\!+\!\frac{\sqrt{3}}{4}q_y\right),
\end{align}
\begin{align}
    \label{eq_d16}
    \mathcal{F}_{BB} &= 16\cos^2 \!\left(\frac{1}{2}q_x\right) \!+\! 8 \sin^2 \! \left(\frac{1}{2}q_x\right),
    \end{align}
\begin{align}
    \label{eq_d17}
    \mathcal{F}_{CC} &= 16\cos^2 \!\left(\frac{1}{4}q_x\!-\!\frac{\sqrt{3}}{4}q_y\right) \!+\! 8 \sin^2 \! \left(\frac{1}{4}q_x\!-\!\frac{\sqrt{3}}{4}q_y\right),
    \end{align}
\begin{align}
    \label{eq_d18}
    \mathcal{F}_{AB} & =\mathcal{F}_{BA}  = 4\cos \!\left(\frac{1}{4}q_x\!+\!\frac{\sqrt{3}}{4}q_y\right) \cos \!\left(\frac{1}{2}q_x\right) \nonumber \\ &\qquad \quad \ \ +\! 4 \sin \!\left(\frac{1}{4}q_x\!+\!\frac{\sqrt{3}}{4}q_y\right)\sin \! \left(\frac{1}{2}q_x\right),
    \end{align}
\begin{align}
    \label{eq_d19}
    \mathcal{F}_{BC} & =\mathcal{F}_{CB}  = 4\cos \!\left(\frac{1}{4}q_x\!-\!\frac{\sqrt{3}}{4}q_y\right) \cos \!\left(\frac{1}{2}q_x\right) \nonumber \\ &\qquad \quad \ \ +\! 4 \sin \!\left(\frac{1}{4}q_x\!-\!\frac{\sqrt{3}}{4}q_y\right)\sin \! \left(\frac{1}{2}q_x\right),
    \end{align}
\begin{align}
    \label{eq_d20}
    \mathcal{F}_{AC} & =\mathcal{F}_{CA}  = 4\cos \!\left(\frac{1}{4}q_x\!+\!\frac{\sqrt{3}}{4}q_y\right) \cos \!\left(\frac{1}{4}q_x\!-\!\frac{\sqrt{3}}{4}q_y\right) \nonumber \\ &\qquad \quad \ \  -\! 4 \sin \!\left(\frac{1}{4}q_x\!+\!\frac{\sqrt{3}}{4}q_y\right)\sin \!\left(\frac{1}{4}q_x\!-\!\frac{\sqrt{3}}{4}q_y\right).
\end{align}

Finally, we evaluate Eq.~\eqref{eq_d14} numerically, to obtain the spin-lattice relaxation rate as expected for the Sb NMR relaxation rate. Figure~\ref{fig_appendix_D} shows that the form factors do not alter the existence of the Hebel-Slichter peak in the kagome lattice, only small quantitative deviations are visible due the different weighting of the momentum structure of the susceptibility in Eq.~\eqref{eq_d14}. This means that the NMR measurements for the unconventional kagome superconductors with $d$-wave pairing symmetry are expected to feature the Hebel-Slichter peak, both on vanadium atoms and on antimony atoms.

\bibliography{Kagome}
\end{document}